\documentclass[11pt,a4paper,notitlepage]{article}
\usepackage[square,numbers]{natbib}
\usepackage{graphicx}
\usepackage{amsmath}
\usepackage{mathptmx}
\usepackage{hyperref}
\hypersetup{pdfborder=0 0 0}
\hypersetup{breaklinks=false}
\hypersetup{colorlinks=true}
\hypersetup{urlcolor=blue}
\hypersetup{linkcolor=blue}
\begin{document}
\newcommand{\mnras}{Mon.\ Not.\ R.\ Astron.\ Soc.}
\newcommand{\apj}{Astrophys. J.}
\newcommand{\pasp}{Publ.\ Astron.\ Soc.\ Pacific}
\newcommand{\aap}{Astron.\ Astrophys.}
\newcommand{\aapr}{A\&AR}

\title{Interpreting radiative efficiency in radio-loud AGN}
\author{M.J. Hardcastle}
\maketitle
\begin{abstract}
  Radiative efficiency in radio-loud AGN is governed by the scaled
  accretion rate on to the central black hole rather than directly by
  the type of accreted matter; while it correlates with real physical
  differences, it does not give us unambiguous information about
  particular objects.
\end{abstract}

Active galactic nuclei (AGN) selected by observations in the optical,
infrared (IR) or X-ray are almost always `radiatively efficient': they
show evidence of a luminous accretion disc, which produces optical and
UV emission directly, photoionizes ambient material to produce broad
and narrow emission-line regions, and drives strong X-ray emission
from a `corona' above the disc. Their radiation may be absorbed and
re-radiated in the mid-IR by a warm, dusty structure referred to as
the `torus'. The diversity of observed properties in these AGN is well
explained on the whole by a combination of intrinsic luminosity
differences and orientation-dependent obscuration
\cite{Urry+Padovani95}.

AGN selected in radio observations can behave very differently. The
evidence is overwhelming that these `radio-loud' AGN (RLAGN) can operate in
both radiatively efficient and inefficient modes \citep[see e.g.][and
  refs
  therein]{Hine+Longair79,Laing+94,Hardcastle+06-2}.
The radiatively inefficient (RI) objects --- known variously as
low-excitation radio galaxies, weak-line radio galaxies
or `jet-mode' objects --- have active jets, which drive the
radio emission, but no evidence for a luminous accretion disc, torus,
corona, or accretion-driven emission lines; by
contrast, radiatively efficient (RE) radio-loud objects ---
high-excitation radio galaxies, narrow-line radio galaxies, broad-line radio galaxies, radio-loud quasars, or
`quasar-mode' objects --- behave like conventional AGN with the
addition of a jet. Nuclear emission in RI objects comes from the jet
only. In terms of their number, RI objects are the
dominant population by a very large factor in the local universe
\cite{Best+Heckman12} and so it is important to understand
their nature and their effects on their environments.

How does this RE/RI dichotomy come about? A long-standing proposal
\cite{Ghisellini+Celotti01,Merloni+Heinz08} is that its basic driver
is the Eddington-scaled accretion rate: objects whose total radiative
luminosity and jet power fall below a few per cent of the estimated
Eddington rate, $L_{\rm Edd} = \frac{4\pi GM_{\rm BH}
  cm_p}{\sigma_{\rm T}}$, are RI, while objects above this limit are
capable of sustaining a conventional RE accretion disc. This is as
expected from theory \citep{Narayan+Yi95}, though the exact threshold
expected depends on details of the accretion disc model. This picture
has recently received strong observational support from large-sample
studies. \cite{Best+Heckman12} estimated total radiative luminosity
from the optical emission-line luminosity of a large sample of RLAGN
drawn from SDSS, NVSS and FIRST, and found a clear division between
RI and RE objects in Eddington-scaled accretion rate, while studies
using X-ray and mid-IR emission as proxies of radiative power instead
\cite{Mingo+14,Gurkan+15} have come to very similar conclusions,
albeit with smaller samples. The basic picture is that all RLAGN
have accretion flows that drive jets of varying powers, but only a
small minority have accretion rates high enough (relative to
Eddington) to also power a RE disc. In RI objects, the estimated
kinetic jet power may exceed upper limits on the radiative luminosity
from the accretion flow by many orders of magnitude.

Ten years ago my collaborators and I \cite{Hardcastle+07-2} proposed
that there was a one-to-one relationship between the two accretion
modes and the two obvious sources of fuel available to AGN in general:
RE sources would be fuelled by cold gas (which could be carried into
the mostly elliptical hosts of RLAGN by mergers, or produced by
cooling) and RI sources by hot gas (directly from the hot phase of the
intergalactic or intracluster medium). A consequence would be that RI objects
are capable of participating in the feedback loop between AGN and hot
gas in cluster centres, while RE sources, which get their fuel from
other sources, may input far more energy into their hot-gas
environments than is required to offset cooling.

This `hot-mode/cold-mode' picture, though widely adopted, has partly
been superseded by our improved understanding over the past decade. If
the Eddington-scaled accretion rate controls the radiative efficiency,
there is no reason why a RI source cannot be produced by accreting at
a low level from a cold gas reservoir, or why rapid accretion from the
hot phase (now widely thought \cite{Gaspari+13} to be mediated by cooling-instability
driven `chaotic accretion' rather than Bondi accretion) should not be capable of sustaining high
accretion rates. Indeed, we can point to individual AGN, such as NGC
1275 in the Perseus cluster, that clearly show signatures of RE
accretion from cold material while also being (presumably)
fundamentally regulated by cooling from the hot phase. Thus at first
sight one might imagine that the `hot-mode/cold-mode' model has no
further role to play. But it is surprising, given these facts, that it
has proved so {\it successful} in predicting or incorporating the
known observational differences between RE and RI objects at low $z$.
Observationally we see that RI objects in general tend to lie in rich
environments, prefer massive quiescent galaxies, and have little star
formation, all as expected if they are fuelled by hot gas; RE objects
tend to lie in poorer environments, prefer smaller galaxies often with
major mergers, and can be strongly star-forming, as would be expected
if there were a copious cold gas supply.

Why does the model make such good predictions if its basic premise is
wrong? I suggest that the best explanation is that there is a strong
{\it association} between the conditions needed to make a RI system
and accretion from the hot phase. The RI/RE threshold can be recast as
a ratio ${\dot M}/M_{\rm BH}$ between the accretion rate and the black
hole mass; for the elliptical hosts of RLAGN $M_{\rm BH}$ scales well
with galaxy mass. Galaxies that are the dominant (most massive)
systems in a group or cluster environment, for a given jet power, will
be more likely to be RI than galaxies in the field both because they
will have high $M_{\rm BH}$ and because the main source of fuel will
be cooling at relatively low $\dot M$ from the hot phase;
merger-driven accretion of large quantities of cold gas in these
systems is disfavoured because of ram-pressure stripping or
strangulation of satellite galaxies in the hot-gas environment. By
contrast, powerful RE AGN, where high accretion rates, close to
Eddington, are required, can most easily be produced in systems with
high $\dot M$ and relatively low $M_{\rm BH}$ and are expected to
favour lower-mass galaxies in field environments. All that is required
for this preference to give rise to significant differences between
the two populations in large samples (as illustrated in Figure 1) is
for there to be a wide range in $\dot M/M_{\rm BH}$ and in RLAGN
environment, both of which we know to be true for powerful sources.
Thus, although the fundamental basis of the model of
\cite{Hardcastle+07-2} has changed, the predictions {\it on a sample
  basis} can remain correct.

\begin{figure}
  \includegraphics[width=1.0\linewidth]{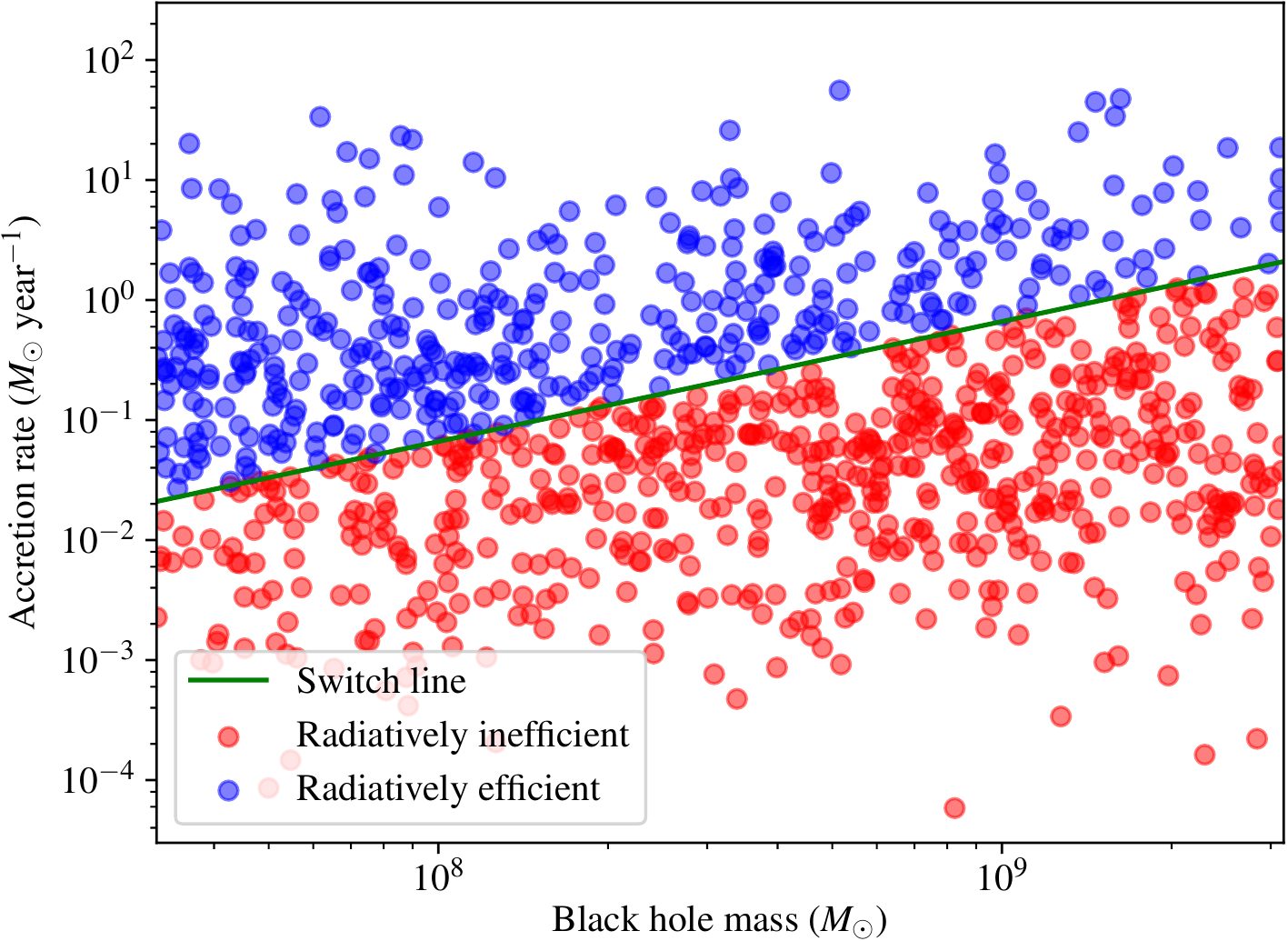}
\caption{Black hole mass and accretion rate in a Monte Carlo
  simulation of a toy model in which black hole mass is uniformly
  distributed in log space between $10^{7.5}$ and $10^{9.5} M_\odot$
  while intrinsic accretion rate, which is unrelated to $M_{\rm BH}$,
  has a lognormal distribution centred on $0.1 M_\odot$ year$^{-1}$
  with a dispersion of 1 dex. The RE/RI division is a straight line in
  this plot corresponding to 3\% of the Eddington rate (assuming 10\%
  conversion efficiency of accreted matter). Even in this toy model,
  RE sources on average have black hole masses a factor 3 less than RI
  sources and available (radiative/kinetic) luminosities, accounting
  for the Eddington limit, 1.2 decades higher, although there is
  essentially no difference between RE and RI sources close to the
  green line. In reality, the accretion rate is likely not
  uncorrelated with the black hole mass, but the details of this
  relationship need to be fleshed out observationally.}
\end{figure}

There are two important consequences of this revised picture. Firstly,
unlike in the `hot-mode/cold-mode' model, the RI/RE observational
classification of any given RLAGN does not {\it necessarily} tell us
anything very fundamental about the source. If a source is close (in
terms of $\dot M/M_{\rm BH}$) to the boundary between the two classes,
it may change type repeatedly over its lifetime as a result of
comparatively small changes in the accretion rate; there is a great
deal of evidence that RLAGN and AGN in general can undergo
large-amplitude long-timescale variation of the accretion power.
Treating the current RI/RE status of a RLAGN as though it were an
immutable property of the system over its lifetime is incorrect and
may lead to other errors. More generally, it is never safe to
conclude that a {\it particular object}'s fuel source is hot gas or
cold gas based on its RI/RE classification; the model predicts a
statistical preference that may be over-ridden by other factors.
Secondly, the results on the cosmological evolution of the two
populations now becoming available \citep{Best+Heckman12} need to be
interpreted with caution, because it is not clear that the
associations between fuel source and accretion mode will operate in
{\it quantitatively} the same way in the earlier Universe; for
example, the evolution of black-hole masses will mean that sources are
more likely to be classed as RE for the same physical accretion rate
$\dot M$ in the earlier Universe, so environments of similar richness
might be expected to produce more RE sources.

In the picture presented here, RE/RI classification is a proxy of a
specific quantity (accretion rate per unit black hole mass) that may
not be particularly relevant to the purpose of a given study. Dividing
observationally generated samples by their radiative efficiency certainly
captures some real physical differences, but other methods of
classifying them might provide more understanding. The ideal situation
would be one in which we had direct estimates of the physically
relevant quantities (jet kinetic power, radiative power, black hole
mass...) and could divide up our samples according to these rather
than simply according to the RI/RE class, but these parameters,
particularly jet power which is fundamental to all studies of RLAGN,
are hard to estimate uniformly and reliably. Developing techniques to
extract jet power from radio observations using Bayesian inference
based on detailed models of the evolution of radio sources
\cite{Turner+Shabala15,Hardcastle18}, and so moving away from the use
of one well-studied but fundamentally limited AGN property, will be
one of the key challenges for RLAGN science in the era of the SKA and
its precursors.

Martin Hardcastle is at the Centre for Astrophysics Research,
University of Hertfordshire, College Lane, Hatfield AL10 9AB, UK.
e-mail: m.j.hardcastle@herts.ac.uk

\bibliographystyle{naturemag}
\bibliography{../bib/mjh,../bib/cards}

\end{document}